\documentclass[12pt]{article}

\usepackage{PRIMEarxiv}

\usepackage[utf8]{inputenc} 
\usepackage[T1]{fontenc}   
\usepackage{hyperref}       
\usepackage{url}            
\usepackage{booktabs}       
\usepackage{amsfonts}      
\usepackage{nicefrac}       
\usepackage{microtype}      
\usepackage{lipsum}
\usepackage{fancyhdr}       
\usepackage{graphicx}       
\usepackage{amsmath}
\usepackage{amssymb}
\usepackage{booktabs}
\usepackage{multirow}
\usepackage{appendix}
\usepackage{array}
\usepackage{subfigure}
\usepackage{caption}

\title{Neuromorphic Incremental on-chip Learning with Hebbian Weight Consolidation
}

\author{
  Zifan Ning\\
  Institute of Automation \\
  Chinese Academy of Sciences\\
  Beijing, China\\
  \texttt{isstillmessup@gmail.com} \\
   \And
  Chaojin Chen  \\
  The Third Affiliated Hospital of \\
  Sun Yat-sen University \\
  Guangzhou\\
  \texttt{chenchj28@mail.sysu.edu.cn} \\
   \And
  Xiang Cheng, Wangzi Yao, Tielin Zhang, Bo Xu  \\
  Institute of Automation \\
  Chinese Academy of Sciences\\
  Beijing, China\\
  \texttt{\{chengxiang2018, yaowangzi2023, tielin.zhang, xubo\}@ia.ac.cn} \\
}

\begin{document}
\maketitle

\begin{abstract}
As next-generation implantable brain-machine interfaces become pervasive on edge device, incrementally learning new tasks in bio-plasticity ways is urgently demanded for Neuromorphic chips. Due to the inherent characteristics of its structure, spiking neural networks are naturally well-suited for BMI-chips. Here we propose Hebbian Weight Consolidation, as well as an on-chip learning framework. HWC selectively masks synapse modifications for previous tasks, retaining them to store new knowledge from subsequent tasks while preserving the old knowledge. Leveraging the bio-plasticity of dendritic spines, the intrinsic self-organizing nature of Hebbian Weight Consolidation aligns naturally with the incremental learning paradigm, facilitating robust learning outcomes. By reading out spikes layer by layer and performing back-propagation on the external micro-controller unit, MLoC can efficiently accomplish on-chip learning. Experiments show that our HWC algorithm up to 23.19\% outperforms lower bound that without incremental learning algorithm, particularly in more challenging monkey behavior decoding scenarios. Taking into account on-chip computing on Synsense\footnote{https://www.synsense.ai/} Speck 2e chip, our proposed algorithm exhibits an improvement of 11.06\%. This study demonstrates the feasibility of employing incremental learning for high-performance neural signal decoding in next-generation brain-machine interfaces.
\end{abstract}

\section{Introduction}

\begin{figure}[ht]
    \centering
    \includegraphics[width=0.5\columnwidth]{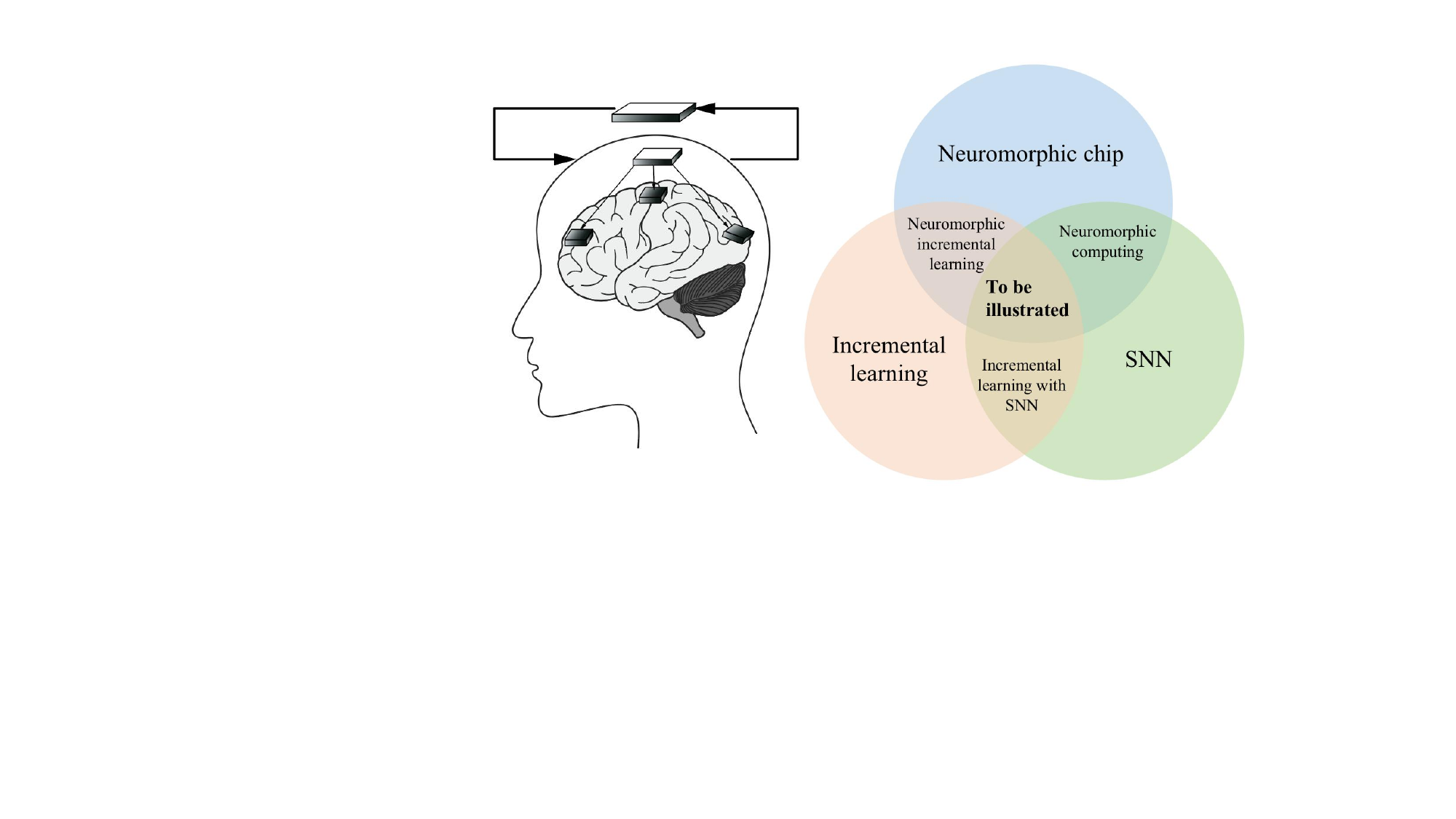} % Reduce the figure size so that it is slightly narrower than the column. Don't use precise values for figure width.This setup will avoid overfull boxes.
    \caption{Incremental on-chip learning in BMI}
    \label{fig_intro}
\end{figure}

The rapid development of Brain-Machine Interface (BMI) \cite{liu2020neural} has made it possible to decode how populations of cortical neurons flexibly execute different functions. \cite{gallego2020long} However, as the number of recording electrodes exponentially increases, the signal processing capabilities of BMI have become a critical issue that needs to be addressed. Neuromorphic computing \cite{ma2022neuromorphic} has emerged as the optimal solution to tackle this challenge. Compared to traditional artificial neural network, spiking neural network (SNN) \cite{zhang2022recent} is better suited to the pulsatile feature of neurons, \cite{wang2023complex} and their high fault tolerance enables robust resistance to strong noise interference in spike coding. \cite{cheng2023meta, jiang2023origin} The incremental learning paradigm trained with SNN accurately captures the dynamic changes in spike coding, leveraging its biological plasticity to faithfully represent the brain's encoding process. Neuromorphic chip, unlike conventional von Neumann chips, employs a working mechanism that closely resembles neurons and synapses, enabling more accurate simulation and replication of the structure and functionality of the human-brain neural systems with extremely low power consumption. \cite{roy2019towards} (See Figure \ref{fig_intro}.)

Incremental learning \cite{singh2021rectification}, as an exceptionally effective algorithm for capturing the dynamic changes in the encoding of neuron population distribution, has been a rapidly developing research direction in the past few years, with numerous new algorithms and frameworks proposed to address the challenges of catastrophic forgetting. \cite{wu2019large} Hebbian learning \cite{tang2023neuro}, a naturally-adaptive learning algorithm for sparse encoding structures in SNNs, is a well-known principle in neuroscience that describes the process by which synaptic connections between neurons are strengthened or weakened based on their activity. \cite{jia2023explaining, zhang2022multi} Since the Hopfield network was first proposed,\cite{krotov2023new} Hebbian learning has been widely studied in the field of neuroscience, and facilitates continuous learning and memory by iteratively adjusting the inter-neuronal connection weights. Neuromorphic incremental learning algorithms based on Hebbian learning \cite{mishra2023survey} exhibit improved accuracy and robustness in handling large-scale data and complex tasks, while also more faithfully emulating the learning and memory mechanisms of the brain.

Neuromorphic computing, where non-volatile technologies \cite{chakraborty2020pathways} have drawn considerable attention for their potential to emulate the functionality of biological synapses in BMI, exhibiting two critical features: synaptic efficacy and synaptic plasticity. \cite{prezioso2018spike} Despite the promising prospects of in-situ computing and synaptic learning for neuromorphic computing, the process of writing to non-volatile devices typically consumes significant resources. The inherent stochasticity of such devices can also lead to unreliable write operations, necessitating costly verification schemes. The high demand for customized chip architectures also entails significant costs. Therefore an urgent need exists for the development of a more simplified on-chip learning solution \cite{wu2021atomically}.

\begin{figure}[t]
    \centering
    \includegraphics[width=\columnwidth]{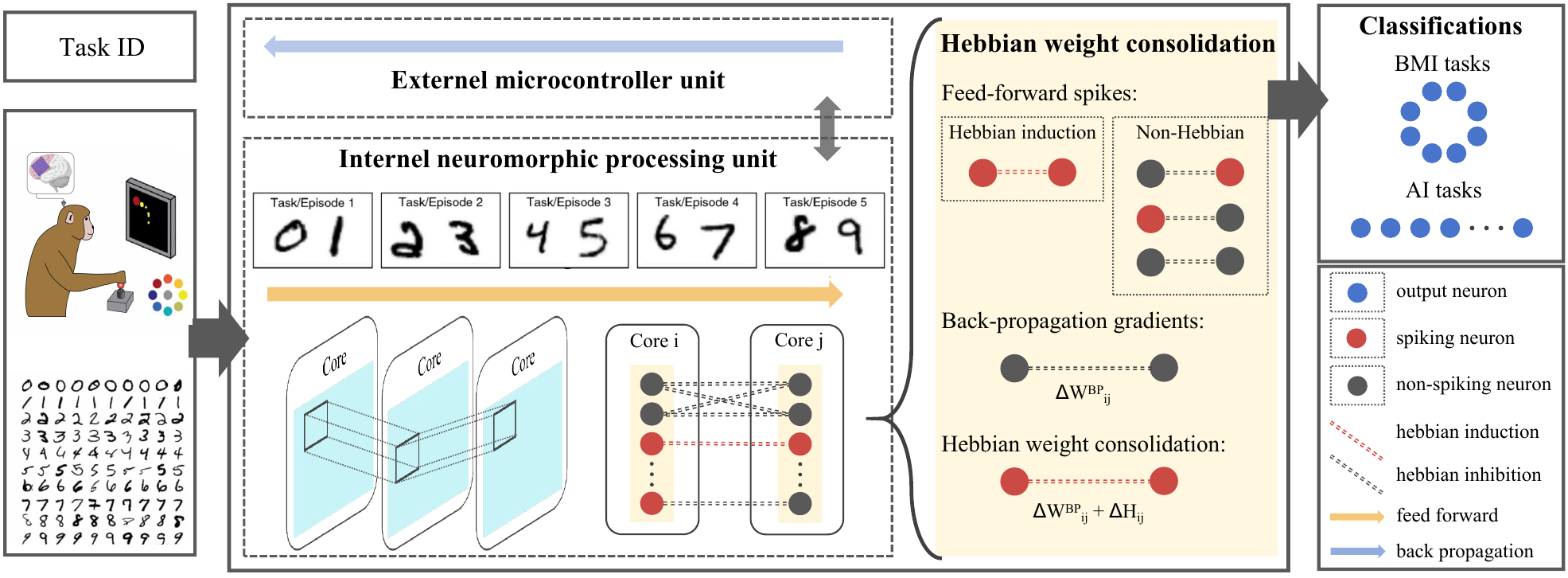} % Reduce the figure size so that it is slightly narrower than the column. Don't use precise values for figure width.This setup will avoid overfull boxes.
    \caption{Mentor-learner on-chip learning framework with Hebbian Weight Consolidation (HWC), where The HWC algorithm performs the forward propagation process on the internal neuromorphic processing unit and sequentially reads out spike sequences core by core. These spike sequences are then transmitted to the external micro-controller unit for back-propagation. After completing one epoch, the updated configurations are sent back to the internal neuromorphic processing unit for the next time step's forward propagation.}
    \label{fig_framework}
\end{figure}

However, currently there is no comprehensive work that integrates neuromorphic chip architecture to implement an incremental learning algorithm based on Hebbian learning in SNN. To address the issue mentioned above, we propose a biologically inspired, incrementally learning algorithm based on SNN, accompanied by a novel neuromorphic mentor-learner on-chip learning framework. Our algorithm outperforms state-of-the-art (SoTA) incremental learning algorithms, particularly in more challenging scenarios, as shown in Figure \ref{fig_framework}. Notably, these algorithms have demonstrated excellent performance when implemented on neuromorphic chips. In summary, our contributions are as follows:

\begin{itemize}
    \item We introduce a biologically plausible SNN-based incremental learning algorithm called Hebbian weight consolidation (HWC), and further illustrate a more hardware-friendly version HWC.
    \item Moreover, we propose a neuromorphic on-chip learning framework based on on-off-chip interaction, called mentor-learner on-chip (MLoC) framework.
    \item We conduct experiments utilizing the split-MNIST, split-DVS10 gesture, CIFAR-100 and monkey cross-day behavior electrocorticography (ECoG) decoding experimental paradigms, and integrate the HWC algorithm and other SoTA incremental learning algorithms onto the Speck Neuromorphic chip such as Elastic weight consolidation (EWC) \cite{kirkpatrick2017overcoming}, Learning without forgetting \cite{li2017learning}, and Synaptic intelligence (SI) \cite{zenke2017continual}.
    \item To the best of our knowledge, We are the first to lie in the proposal of BMI based on neuromorphic on-chip incremental learning.
\end{itemize}

\section{Related Works}

\noindent{\textbf{Invasive Brain-machine interface: }Invasive BMI has successfully been utilized to achieve precise control of prosthetic and external device, allowing individuals with disabilities to restore their limb functions. \cite{chiavazza2023low, chakraborty2020pathways, shanechi2019brain} Additionally, invasive BMI has been applied in the treatment of neurological disorders such as epilepsy and Parkinson's disease, where implanted electrodes are used to stimulate or record brain activity, thus improving disease symptoms. Furthermore, invasive BMI serves as a crucial tool for neuroscience research, enabling the recording and analysis of neural activity to gain deeper insights into the functionality and workings of neural circuits in the brain. In the field of behavior decoding and motor control, researchers have been able to accurately predict and interpret individuals' intentions and actions by decoding brain signals. \cite{gallego2020long, feulner2022small} Challenges that currently need to be addressed include the precision and stability of signal decoding, the portability and long-term usability of systems, among others, in order to achieve broader clinical applications.}

\noindent{\textbf{Incremental learning: }Incremental learning has continued to be an active research area, with many new approaches proposed to address the catastrophic forgetting challenge from non-stationary data. Gido \cite{van2022three} describes three fundamental types of incremental learning: task, domain and class-incremental learning to help address this area. Based on the principles of their implementation, recent incremental learning algorithms can be categorized into three main classes: regularization-based \cite{kirkpatrick2017overcoming, zenke2017continual}, replay-based \cite{rebuffi2017icarl, van2020brain}, and dynamic structure-based \cite{xue2022meta, wang2022learning, simon2022generalizing} methods. One recent incremental learning trend is to incorporate meta-learning into incremental learning, which allows models to quickly adapt to new tasks with limited data. \cite{zhuang2023gkeal, wang2022meta}. Another trend is to use generative models for incremental learning. \cite{zhou2022forward, zhu2022self} These recent works have demonstrated the potential of incorporating meta-learning and generative models into incremental learning.
}

\noindent{\textbf{Neuromorphic on-chip computing: }There has been an increasing interest in neuromorphic on-chip learning due to its potential for efficient and low-power machine learning. \cite{ma2022neuromorphic} Various neuromorphic architectures have been proposed, such as the Tianji \cite{shi2015development} and Loihi \cite{davies2018loihi}, that can be implemented on neuromorphic hardware to develop hardware platforms capable of supporting on-chip learning. In addition, exploring new algorithms and training methods that can be implemented on neuromorphic hardware has become a popular research area among scholars. \cite{roy2019towards,sandamirskaya2022neuromorphic,imam2020rapid} In the realm of incremental learning on neuromorphic chips, some have leveraged synaptic plasticity-based methods, namely long-term potentiation (LTP) \cite{sumi2020mechanism} and short-term potentiation \cite{miranda2020modeling} (STP). Additionally, some researchers have advocated for methods that capitalize on the dynamic behavior of neurons and implement incremental learning by modulating the dynamic behavior of neurons. \cite{zhu2021self,hu2021distilling}}

\section{Methodology}

\subsection{Hebbian weight consolidation (HWC)}

To mitigate the catastrophic forgetting problem, we introduce a low energy-consumption consolidation mechanism inspired by the Hebbian theory which establishes relationship between synaptic plasticity $\Delta {W}^T_{i,j}$ with neuronal activity of pre- and post-synaptic, as shown in Formula \ref{traditional_hebb}.
\begin{equation}
\label{traditional_hebb}
    \Delta {W}^\tau_{i,j} = \Delta W^{\tau,BP}_{i,j} \cdot P^\tau_{i,j},
\end{equation}
where $i, j$ represent pre- and post-synaptic of neuron respectively. $\tau$ is the current task. $P$ represents the potential, which could be calculated as Formula \ref{soft_mask}:
\begin{equation}
\label{soft_mask}
    P^{\tau=t}_{i,j} = 1-\max(H^{1}_{i,j}, H^{2}_{i,j}, H^{3}_{i,j}, \cdots H^{t-1}_{i,j}),
\end{equation}
where $t$ represents the number of total tasks of the increment scenario. Here we define $H$ as the Hebbian information, which can be calculated according to the firing rate $f$, as shown in Formula \ref{hebb_info}:
\begin{equation}
    \label{hebb_info}
    H^\tau_{i,j} = \frac{1}{N}\sum_{i,j} f_i \cdot f_j.
\end{equation}
Though Hebbian theory is often used as a learning algorithm for AI models, \cite{zhang2021self} we adopt it to calculate the potential of synaptic modification which furtherly determines whether synaptic modification should be restricted in the following tasks of an incremental learning procedure, according to the coincident firing frequency of synapses. This potential, which is based on neuronal activity, can indicate the importance of certain synapses in different tasks of incremental learning. During the learning procedure of a task, the firing rate of neurons or the coincident firing frequency of synapses are recorded. Before training a new task, the information from previous tasks is integrated and compared to a predefined threshold, generating a 0-1 mask which disables the further modification of certain synapses and thus consolidation the former modification. 

Moreover, we provide a more hardware-friendly version to calculate the potential between pre- and post-synaptic, as shown in Formula \ref{hard_mask}:
\begin{align}
\label{hard_mask}
    P^{\tau=t}_{i,j} &= 1 - g(\max(H^{1}_{i,j}, H^{2}_{i,j}, H^{3}_{i,j}, \cdots H^{t-1}_{i,j})), \\
    g(x) &= \begin{cases}
        0, \quad x \leq threshold \\
        1, \quad x > threshold.
    \end{cases}
\end{align}
In the hard masking strategy, binary potentials are better suited for bitwise operations on neuromorphic chips, thus improving hardware computational efficiency. By selectively masking the modification of synapses that are crucial for previous tasks, new knowledge from subsequent tasks can be stored in retained synapses, while the old knowledge from these tasks is consolidated and preserved. Compared to other SoTA algorithms, HWC only needs to calculate the local Hebbian information and store the importance matrix for tasks already trained, thus possess low energy consumption and memory space, making it friendly to hardware implement.

HWC is inspired by Hebbian learning \cite{zhao2020glsnn} and is built on the fundamental premise of cognitive class chips. Considering the limited model capacity and the vast amount of new information to learn, HWC recognizes the necessity of selectively preserving or deleting knowledge. HWC tackles the issue of catastrophic forgetting in incremental learning through two primary aspects. Firstly, during the back-propagation process, HWC employs synaptic plasticity as a fundamental principle, allowing for the adjustment of synaptic strengths based on the Task ID of the input pattern. This plasticity allows the model to function as a task incremental learning algorithm in AI tasks, which is the primary learning mode of HWC. Besides, during the forward propagation process, HWC incorporates Hebbian learning, which is based on the inter-neuronal associations. This learning rule enables the HWC algorithm to be applicable to time-coding-based BMI tasks, particularly those primarily focused on domain incremental learning scenarios.

\subsection{Mentor-learner on-chip learning framework}

\begin{figure}[t]
    \centering
    \includegraphics[width=0.7\columnwidth]{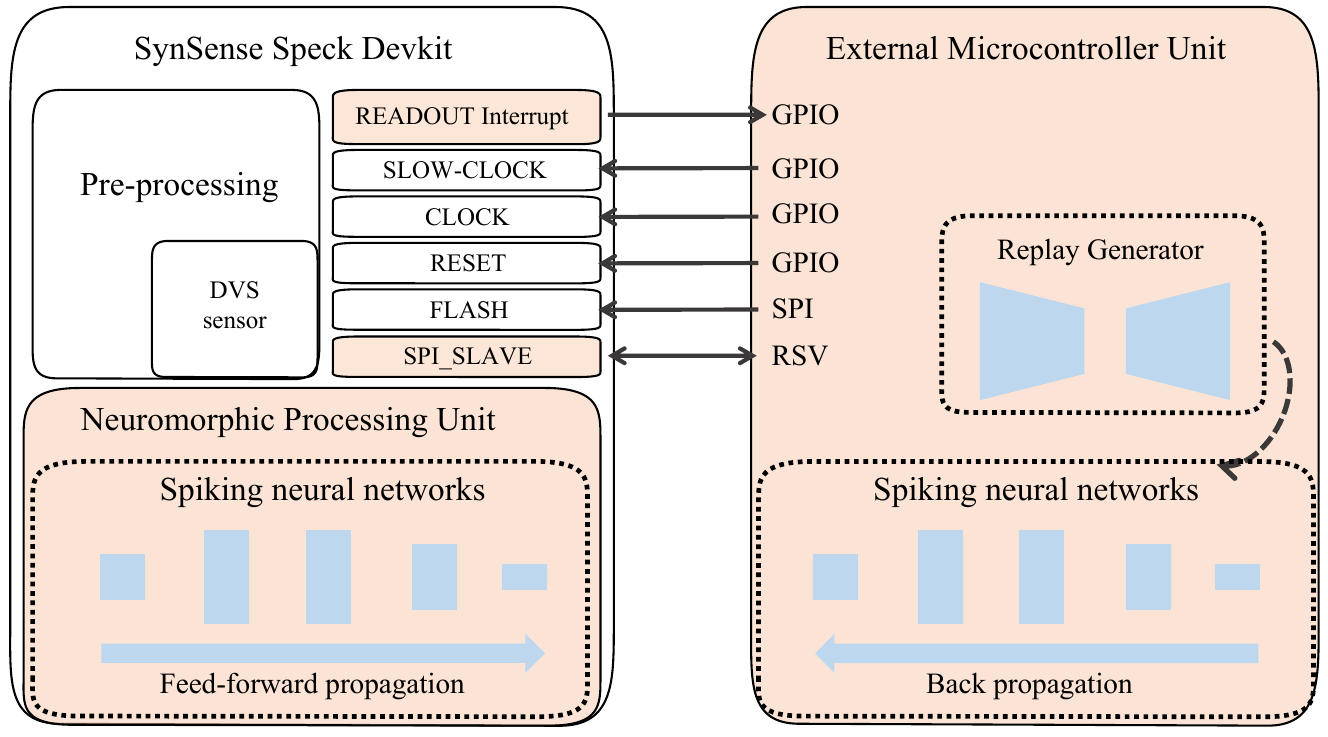}
    \caption{MLoC framework on Speck chip.}
    \label{fig_colearning}
\end{figure}

Here we propose a mentor-learner on-chip learning framework (MLoC) based on the offchip(mentor)-onchip(learner) interaction, as shown in Figure \ref{fig_colearning}.

The learner part is constructed using an internal neuromorphic processing unit (INPU), which can theoretically be any type of neuromorphic processing unit. The mentor part is an external microcontroller unit (EMU), which can be any architecture of neural network processor capable of performing tensor calculations. It can be combined with the INPU in the form of a daughterboard. In this study, we choose to utilize a GPU as the EMU to enhance computational efficiency. MLoC framework achieves on-chip learning by reading out the on-chip neural spikes layer by layer through an EMU, and by training the same structured neural network while providing feedback to INPU. Because of the structural consistency of the external neural network, it can accommodate more advanced configurations, such as dropout and normbatch. As a result, the MLoC framework can achieve superior performance compared to hardware that can only perform on-chip weight updates.

On the Speck chip, the access and modification of on-chip registers and SRAM values are performed by an external processor through the SPI-slave interface. The chip's array readout, which is performed layer by layer, is stored in registers. Once the configuration is finalized and computation starts, the external processor can monitor the Interrupt signal's changing state to wake up the system and retrieve the calculated results from the READOUT. After the back-propagation process is completed, a new set of configurations is sent back to the Speck chip through the SPI interface to complete a training iteration.

To this end, this paper proposes the on-off interaction cross-entropy loss function, which combines external neural networks, internal neural networks, and the hardware simulation environment for internal neural networks on external processor, as shown in Formula \ref{cceloss}.
\begin{equation}
\label{cceloss}
    L=\lambda H(y,\hat{y}_{ENN})+\mu(\alpha H(y,\hat{y}_{INN}) + \beta H(y,\hat{y}_s)),\\
\end{equation}
where $y$ represents the ground truth. $\hat{y}_{ENN}$ represents the prediction of external neural network on external processor. $\hat{y}_{INN}$ represents the prediction of internal neuromorphic network on internal neuromorphic processor. $\hat{y}_s$ represents the prediction of simulated neuromorphic network on external processor. While $\lambda$, $\mu$, $\alpha$ and $\beta$ are four hyper parameters that can be adjusted, where if $\mu$ is set to 0, MLoC will be a fully external training mode, where the model is deployed onto neuromorphic chip after well-trained. In AI tasks primarily focused on image recognition, this setting will not significantly impact training accuracy and can greatly enhance training speed.

According to the interaction logic, frequent on-off-chip interactions during the entire on-chip learning process can potentially result in high power consumption associated with the read and write operations of the RAM. However, subsequent experiments have demonstrated that proposed framework can achieve on-chip incremental learning with extremely low chip power consumption.

\subsection{Evaluation metrics}

As adopted in \cite{zhu2021prototype}, We present the average accuracy and incremental accuracy as performance metrics, evaluated across three distinct runs, to assess our system's performance. Besides, we further adopt incremental accuracy \cite{zhang2023brain} as illustrated in Formula \ref{incre_acc}:
\begin{equation}
    Acc^{\leq c}_{\Delta} = \frac{1}{c} \cdot \sum^c_{\tau=1}Acc^{\tau},
    \label{incre_acc}
\end{equation}
where $Acc_{\Delta}$ is the incremental accuracy. $Acc^{\tau}_{\Delta}$ is the test accuracy for the $\tau$-th class, and $c$ represents the total number of classes of the scenario. The average incremental accuracy represents the average accuracy across all incremental stages, including the initial stage, providing a fair comparison of the overall incremental performance among different methods.

In addition, we employ the following five components to measure the power consumption of the neuromorphic chip: IO power consumption, Power consumption of read and write of ram, Logic operation power consumption, DVS digital power consumption, and DVS analog power consumption.

\section{Experiments}

\subsection{Compared methods}

Due to the limitations of neuromorphic chip architectures, methods that attempt to achieve compatible dynamic network structures by altering or dividing the network are not suitable for implementing on-chip learning in neuromorphic systems. These methods often necessitate more complex customized chip architectures, and their high memory requirements and low computing efficiency make them unsuitable for high real-time neuromorphic experimental scenarios.

To gauge the effectiveness of HWC, we compare it against seven SoTA non-dynamic structured incremental algorithms on SNNs. We consider three widely used regularization-based methods: Elastic Weight Consolidation (EWC) \cite{kirkpatrick2017overcoming}, Learning without Forgetting \cite{li2017learning}, and Synaptic Intelligence (SI) \cite{zenke2017continual}. Another SoTA approach to incremental learning is Context-dependent Gating (XdG) \cite{masse2018alleviating}, which selects a different randomly chosen subset of network nodes for each task to reduce interference between tasks. Furthermore, iCaRL \cite{rebuffi2017icarl} combines classifier and representation learning techniques to enable continuous learning of new classes in a dynamic environment while preserving accuracy for previously learned classes. Moreover, Generative Replay (GR) \cite{kingma2013auto} and Brain-inspired Generative Replay \cite{van2020brain} are two widely used SoTA generative-based incremental learning algorithms. They effectively mitigate the problem of forgetting old tasks by utilizing generative models.

In addition, the Joint represents the training results of non-incremental learning, which is the ideal upper bound we aim to achieve using incremental learning algorithms. The None, on the other hand, represents the results obtained from sequential learning without any incremental learning algorithm, indicating a lower bound that is prone to catastrophic forgetting.

\subsection{Experimental paradigms}

The experimental paradigm is constrained by the fact that the Speck development board supports a maximum of 16-class output, which limits our options for experimentation. 

\noindent{\textbf{Split-MNIST:} The concept of split-MNIST is first introduced in Paper \cite{zenke2017continual}. To evaluate the performance of continuous learning models, the authors divided the MNIST dataset into 5 parts, forming 5 tasks. Here, we utilize the split-MNIST dataset to perform task incremental learning scenario.}

\noindent{\textbf{CIFAR-100:} We further validate the off-chip performance of the HWC algorithm using the CIFAR-100 dataset. \cite{krizhevsky2009learning} CIFAR-100 consists of 100 different classes of color images, with each class having 600 training samples and 100 testing samples. Here, we divide the 100 classes into ten tasks equally and perform incremental training sequentially. Here, we utilize the CIFAR-100 dataset to perform task incremental learning scenario.}

\noindent{\textbf{Split-DVS10 gesture:} We reshape the IBM DVS128 dataset \cite{amir2017low} to a ten-class incremental learning version. Specifically, we removed the action "air guitar" and grouped the remaining ten actions into pairs to form individual tasks. The reshaped DVS128 dataset is serialized into five tasks, following the paradigm of task incremental learning, and used as input for our experimental model. The purpose of this approach is to enable Neuromorphic chips equipped with DVS cameras to incrementally learn new tasks in the DVS object recognition system.}

\noindent{\textbf{Behavior decoding:} Behavior decoding is an eight-classification center-out dataset. We employed long-term implanted microelectrode arrays to record neural activity in the M1 cortex of monkeys, and select seven days of recorded data for the current experimental scenario. In particular, behavior decoding represents a long-term, unbounded domain incremental learning paradigm, making it unsuitable for evaluation using average accuracy. Here, we utilize the time-coding-based behavior decoding dataset to perform domain incremental learning scenario.}

\subsection{Internal neuromorphic processor}

\begin{figure}[t]
    \centering  %居中
    \subfigure[Speck 2E chip]{   %第一张子图
    \begin{minipage}{0.29\columnwidth}
    \centering    %子图居中
    \includegraphics[width=\columnwidth]{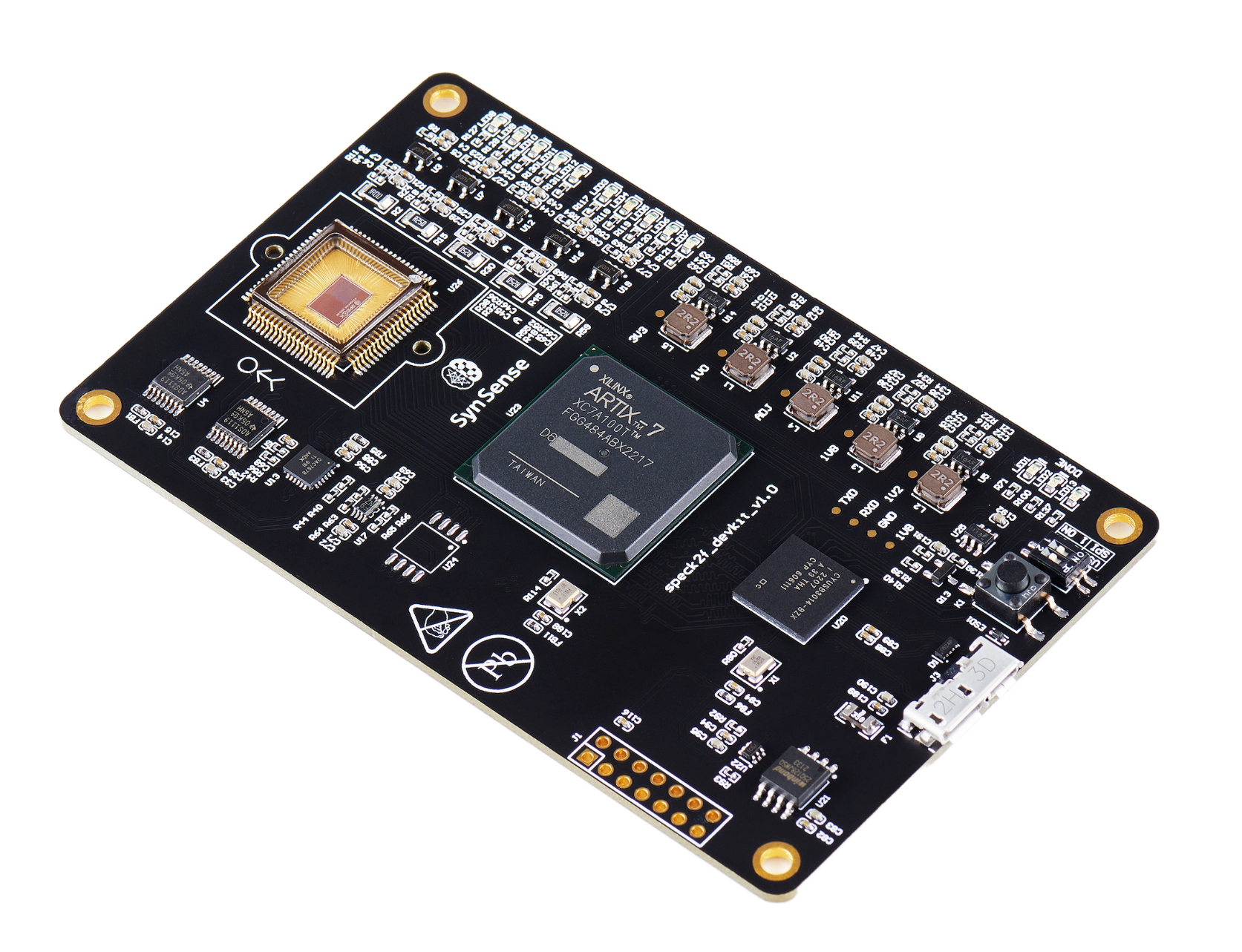}  %以pic.jpg的0.5倍大小输出
    \label{fig_speck1}
    \end{minipage}
    }
    \subfigure[Speck 2E internal architecture]{ %第二张子图
    \begin{minipage}{0.67\columnwidth}
    \centering    %子图居中
    \includegraphics[width=\columnwidth]{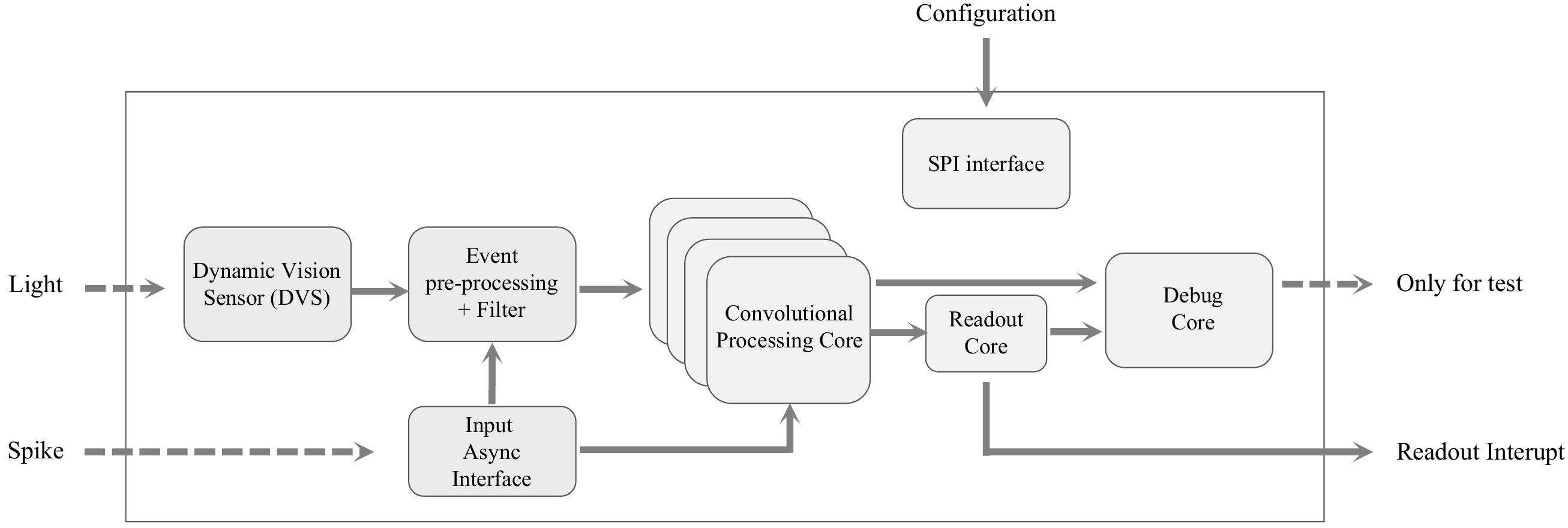}%以pic.jpg的0.5倍大小输出
    \label{fig_speck2}
    \end{minipage}
    }
    \caption{External and internal architecture of the Speck 2E dev-kit}    %大图名称
    \label{dvs_speck}    %图片引用标记
\end{figure}

Here, we use the Speck 2E development board developed by Synsense\footnote{https://www.synsense.ai/}. The appearance of Speck Devkit is shown in Figure \ref{fig_speck1}. Speck is a multi-core single-chip event-driven neural morphology processor chip with a DVS sensor, suitable for real-time mobile and IoT visual applications. It features a configurable Spiking CNN architecture and up to 320,000 neurons, making it well-suited for processing neural morphology data and handling real-time applications.

As shown in Figure \ref{fig_speck2}, Speck is equipped with 9 configurable spiking convolutional computing layers. The layers can each implement a layer of a SCNN neural network, and can be connected to form a user defined network of any size up to the maximum available resources. Layer memory sizes are balanced to provide a flexible balance of resources, with larger or smaller layers, in terms of kernel size or neuron number.

\subsection{Settings}
Due to the constraints imposed by the neural morphological chip architecture and limitations in kernel memory and neuron memory, the selection of SNN structures lacks high flexibility. Each layer's parameters in the neural network need to be carefully calculated.

We opt to avoid using bias in our approach. The inclusion of bias introduces significant overhead on the chip, as it involves non-sparse operations, thereby increasing the computational workload and consequently power consumption, particularly in layers with a large number of neurons. This is primarily because the bias terms are inherently linked to the leakage mechanisms of neurons on the hardware level. The leakage rate of neurons is influenced by an external slow clock.

\subsection{HWC validation experiment}

We first conducted performance comparisons of the HWC algorithm with seven other SoTA algorithms on SNNs across four experimental paradigms in the off-chip environment, as shown in Table \ref{off_incre_acc}. Our algorithm up to 7.50\% outperforms the majority of existing non-dynamic structural incremental learning algorithms in terms of average accuracy across all tested scenarios. Particularly, it demonstrates superior performance on more challenging and complex time-series datasets, such as behavior decoding and DVS gesture datasets.  In the behavior decoding scenario, our method outperforms other SoTA incremental learning algorithms with incremental accuracy improvements up to 8.12\%. Besides, in the incremental DVS128 gesture recognition scenario, the HWC algorithm demonstrates accuracy second only to XdG, 0.76\% - 21.16\% outperforms other SoTA incremental learning algorithms. In more challenging AI task CIFAR100, HWC has also shown excellent performance. After ten tasks of incremental learning, HWC achieved an overall evaluation accuracy of 52.22\%, ranking second only to the brain-inspired replay method.

\begin{table*}[!t]
\renewcommand\arraystretch{1.3}
    \centering
    \caption{Off-chip incremental accuracy of five scenarios on SNNs.}
    \resizebox{\linewidth}{!}{
    \begin{tabular}{c|ccc|ccc|ccc|ccc|c}
         \hline
         \multirow{2}{*}{Models} & \multicolumn{3}{|c|}{Split-MNIST (\%)} & \multicolumn{3}{|c|}{Split-DVS10 (\%)} & \multicolumn{3}{|c}{Behavior decoding (\%)}&\multicolumn{3}{|c|}{CIFAR100 (\%)} &  \multirow{2}{*}{Average (\%)} \\
         \cline{2-13}
          & $Acc^{\leq1}_{\Delta}$ & $Acc^{\leq3}_{\Delta}$ & $Acc^{\leq5}_{\Delta}$ & $Acc^{\leq1}_{\Delta}$ & $Acc^{\leq3}_{\Delta}$ & $Acc^{\leq5}_{\Delta}$ & $Acc^{\leq1}_{\Delta}$ & $Acc^{\leq3}_{\Delta}$ & $Acc^{\leq5}_{\Delta}$ & $Acc^{\leq1}_{\Delta}$ & $Acc^{\leq5}_{\Delta}$ & $Acc^{\leq10}_{\Delta}$  \\
         \hline
         Joint &  & 99.39 &  &  & 92.92 &  & &87.34 &  & & 57.33 & &84.25  \\
         None & 96.31 & 88.72 & 83.73 & 97.33 & 57.93 & 45.91 & 86.50 &58.11 &36.37 &53.19 &34.18 &20.30 & 52.27 \\
         \hline
         iCaRL \cite{rebuffi2017icarl} & 97.32 & 93.53 & 92.53 & 92.54 & 78.33 & 70.19 & 87.23 & 60.11& 45.19 & 55.33 & 49.81& 47.33 & 72.45 \\

         EWC \cite{kirkpatrick2017overcoming} & \textbf{99.95} & 96.36 & 94.11 & 97.63 & 89.16 & 81.61 &87.45 &62.23 &50.74 &56.50 & \textbf{53.31}& 51.94& \textbf{76.74} \\

         SI \cite{zenke2017continual} & 99.76 & 97.35 & \textbf{97.67} & \textbf{97.65} & \textbf{90.48} &79.83  &\textbf{87.55} &61.13 &49.69 & 56.50&51.11 & 49.51 & 76.51 \\

         XdG \cite{masse2018alleviating} & 92.13 & 89.75 & 88.83 & 97.63 & 72.81 & 61.21 &87.02 &61.10 &49.56 & 54.59 & 33.22 & 27.61& 67.96 \\

         LwF \cite{li2017learning} & 94.01 & 93.12 & 90.54 & 98.42 & 71.14 & \textbf{88.46} &87.16 &61.22 &47.50 & 55.13 & 40.41 &38.08 & 72.10 \\

         GR \cite{kingma2013auto} & \textbf{99.95} & \textbf{98.36} & 89.32 & 95.33 & 72.31 & 65.01 &\textbf{87.55} &61.43 &49.11 & 51.20& 41.94 &46.33 & 71.49 \\

         Brain-inspired \cite{van2020brain} & 98.52 & 97.14 & 96.37 & 96.71 & 74.11 & 70.04 &\textbf{87.55} &63.21 &49.67 & \textbf{57.81} & 52.44 & \textbf{53.01} & 74.71 \\
         %&&&&&&&&&&&&&&&\\
         \hline
         
         HWC (proposed) & 96.32 & 94.59 &93.46 & 97.28 & 87.83 & 82.37 &\textbf{87.55} &\textbf{65.73} &\textbf{55.62} &56.83 & 53.29&52.22 & 75.46 \\

         %HWC+GR+SI+LwF & 98.56 & 97.49 &94.01 & &  &  &87.05 &68.42 &60.11 & & & & & & \\
         \hline
    \end{tabular}
    }
    \label{off_incre_acc}
\end{table*}

\subsection{HWC on-chip computing}

In this section, we successfully deployed the mentioned SoTA algorithms and the proposed HWC algorithm using the MLoC framework on the Speck neuromorphic chip for three experimental paradigms: split-MNIST, DVS gesture, and behavior decoding, as shown in Table \ref{on_incre_acc}. Due to the maximum classification output limitations of the neuromorphic chip, our experimental paradigms were limited to scenarios with fewer than 16 classes. The specific experimental results are as follows.

\noindent{\textbf{Non-increment hardware performance: }Speck Neuromorphic chip achieved an accuracy of 95.33\% on the MNIST dataset. In comparison, the architecture proposed by \cite{zhang2018sign}, which utilizes a RRAM crossbar-based neuromorphic computing architecture for MLP, along with an on-chip learning algorithm called SBP, achieved a classification accuracy of 94.5\% on the non-incremental learning task of MNIST. Our method, with a simpler implementation and faster convergence speed, achieved an improvement of approximately 0.8\% in accuracy. Besides, the Speck Neuromorphic chip with on-chip learning achieved a classification accuracy of 91.27\% in DVS128 incre-gesture scenario, and 84.19\% in the behavior decoding non-incremental learning scenario.}

\noindent{\textbf{Incremental accuracy: }For the incremental learning task, the Speck Neuromorphic chip, equipped with the HWC method, demonstrates an average accuracy improvement of 11.37\% in the split-MNIST paradigm compared to non-incremental learning algorithms, 34.75\% in the split-DVS10 paradigm, and 16.93\% in the behavior decoding paradigm, indicates that the HWC algorithm effectively mitigates the problem of catastrophic forgetting in incremental learning.

Compared to other methods, HWC attained the overall accuracy of 80.85\% on the Speck Neuromorphic chip, second only to SI method, 81.06\%. The primary growth contributions arise from the BMI task behavior decoding scenario and the time-coding-based DVS Incremental dataset in AI tasks.}

In the split-MNIST task incremental learning scenario, the best overall performance is achieved by SI, followed by our proposed HWC method, 93.42\%. In the split-DVS10 task incremental learning scenario, our algorithm ranks just below XdG and SI methods, 75.86\%.

In terms of behavior decoding domain-incremental learning scenario, the SNN equipped with the HWC algorithm achieved an incremental within-3-day accuracy of 63.19\%, surpassing the SoTA solutions by 3.88\% to 7.47\%. Besides, incremental accuracy of behavior decoding within-5-day is 55.62\%, surpassing the SoTA solutions by 4.30\% to 11.11\%, as shown in Table \ref{on_incre_acc}.

In time-coding-based incremental training scenarios like split-DVS10 gesture and behavior decoding, the within-1-epoch incremental accuracy indicates that the HWC algorithm may not be the fastest method for learning new knowledge. However, as the incremental training progresses, HWC effectively delays catastrophic forgetting, suggests that the HWC algorithm is suitable for long sequences and multiple incremental tasks on neuromorphic chips, aligning with our intended design for BMI long-term behavioral decoding device.

\begin{table*}[!t]
\renewcommand\arraystretch{1.1}
    \centering
    \caption{On-chip incremental accuracy of three tasks on Speck.}
    \scalebox{0.95}{
    \begin{tabular}{c|ccc|ccc|ccc|c}
         \hline
         \multirow{2}{*}{Models} & \multicolumn{3}{|c|}{Split-MNIST (\%)} & \multicolumn{3}{|c|}{Split-DVS10 (\%)} & \multicolumn{3}{|c|}{Behavior decoding (\%)} & \multirow{2}{*}{Average} \\
         \cline{2-10}
          & $Acc^{\leq1}_{\Delta}$ & $Acc^{\leq3}_{\Delta}$ & $Acc^{\leq5}_{\Delta}$ & $Acc^{\leq1}_{\Delta}$ & $Acc^{\leq3}_{\Delta}$ & $Acc^{\leq5}_{\Delta}$ & $Acc^{\leq1}_{\Delta}$ & $Acc^{\leq3}_{\Delta}$ & $Acc^{\leq5}_{\Delta}$ \\
         \hline

         Joint &  & 95.33 &  &  & 91.27 &  & &84.19 & & 90.26 \\

         None & 94.19 & 88.42 & 82.05 & 97.31 & 55.62 & 41.11 &81.20 &53.19 &35.08 & 69.79 \\
         \hline
         iCaRL \cite{rebuffi2017icarl} & 97.67 & 91.20 & 90.55 & 90.13 & 78.43 & 71.29 & 81.11 & 57.09 & 47.71  & 78.35 \\

         EWC \cite{kirkpatrick2017overcoming} & 97.23 & \textbf{96.14} & 91.71 & 93.66 & 82.19 & 75.70 &83.49 &59.31 &45.22  & 80.59 \\

         SI \cite{zenke2017continual} & 97.43 & 95.19 & \textbf{95.33} & 93.33 & \textbf{85.39}  & 74.66 &84.10 &58.89 &45.29  & \textbf{81.06} \\

         XdG \cite{masse2018alleviating} & 90.56 & 85.45 & 87.01 & 96.46 &73.44  &  59.02 &81.31 &55.72 &43.02  & 74.67 \\

         LwF \cite{li2017learning} & 93.44 & 91.56 & 89.90 &\textbf{97.30} & 70.88 & \textbf{85.56} &82.47 &56.57 &40.90  &78.73  \\

         GR \cite{kingma2013auto} & \textbf{98.31} & 95.70 & 90.63 & 94.99 & 71.30 & 66.41 &83.67 &58.80 &47.61 & 78.60  \\

         Brain-inspired \cite{van2020brain} & 95.55 & 96.13 & 90.21 & 95.50 & 72.33 & 69.00 &83.51 &58.37 &47.11  & 78.62 \\
         \hline
         HWC (proposed) & 95.13 & 93.91 &93.42 & 95.93 & 72.33 & 75.86 & \textbf{85.89} & \textbf{63.19} & \textbf{52.01}  & 80.85 \\
         %HWC+GR+SI+LwF & 97.32 & 94.59 &93.46 & &  &  &87.05 &65.73 &55.62 & \\
         \hline
    \end{tabular}
    
    }
    \label{on_incre_acc}
\end{table*}

\begin{figure*}[t]
    \centering
    \vspace{1pt}
    \includegraphics[width=\columnwidth]{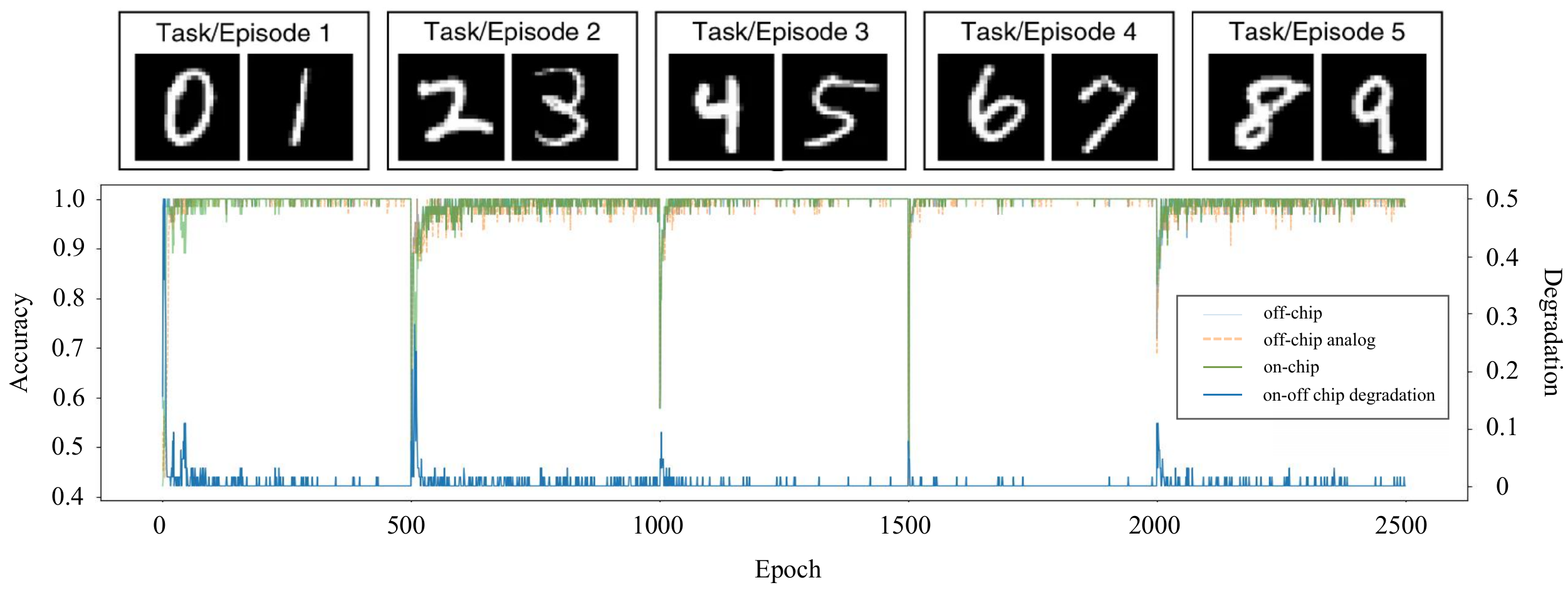} % Reduce the figure size so that it is slightly narrower than the column. Don't use precise values for figure width.This setup will avoid overfull boxes.
    \caption{Training fit of on-off chip learning on Split-MNIST scenario}.
    \label{fig_mnist5cls}
\end{figure*}

\noindent{\textbf{Learning goodness-of-fit: }Moreover, MLoC framework has demonstrated excellent performance across several algorithms, enabling the integration of multiple SoTA algorithms onto the Speck Neuromorphic chip with minimal on-chip accuracy attenuation and high training efficiency. As shown in the Figure \ref{fig_mnist5cls}, during the training process, the on-chip accuracy closely follows the accuracy of the off-chip model with only a few steps of lag. The overall accuracy degradation after well-trained is lower than 5\%}

\subsection{Further analysis}

\noindent{\textbf{On-hardware accuracy degradation: }Despite the use of proposed on-chip learning framework, the on-chip learning algorithm still experiences a certain level of accuracy decay. It is important to note that such decay is amplified as the performance of the original off-chip network decreases. In this section, we focus on analyzing the reasons behind the on-chip accuracy decay.}

In the case of non-incremental joint training, the off-chip accuracy for the DVS gesture recognition scenario is 92.92\% . When the same architecture and experimental data are deployed on the neuromorphic chip, the accuracy drops to 91.27\% , resulting in a degradation of 1.65\%. In the case of one of the well-performing algorithm SI, the on-off neuromorphic chip degradation on behavior decoding within-5-day incremental accuracy is 4.22\%, while for the HWC algorithm, it is only 0.21\%. The behavior decoding scenario exhibits a high decay rate due to a mismatch between the time coding of the dataset and the bandwidth constraints of the Speck chip (SynOps/s), resulting in internal congestion within the chip. However, the HWC algorithm, based on the Hebbian rule, effectively preserves spike information during the training process, thereby reducing the degradation. 

The comparison between on-chip and off-chip accuracy degradation indicates that our proposed algorithm achieves lower accuracy degradation on the neuromorphic chip. This is particularly evident in the behavior decoding paradigm, which relies on time coding as the data representation. These results demonstrate that our algorithm is more neuromorphic-friendly. Moreover, the HWC algorithm, with its biological plasticity, is well-suited to the computational architecture of SNNs and the time coding data representation on the neuromorphic chip.

\begin{figure}[h]
    \centering
    \includegraphics[width=0.7\columnwidth]{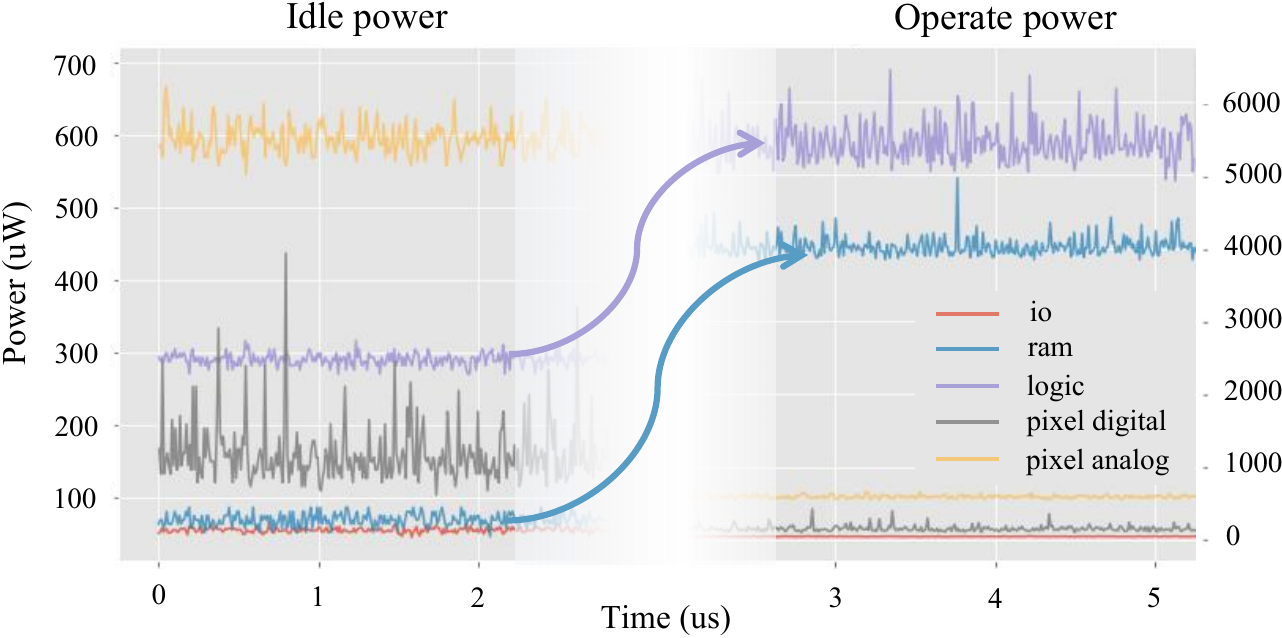} % Reduce the figure size so that it is slightly narrower than the column. Don't use precise values for figure width.This setup will avoid overfull boxes.
    \caption{MLoC power analysis on Speck chip.}
    \label{chip_power}
\end{figure}

\noindent{\textbf{Energy consumption: }As shown in the Figure \ref{chip_power}, when on-chip learning is implemented on the Speck chip with five DynapCNN layers operating at full capacity, only approximately 6000 uW of logic operation power is required for on-chip incremental learning. Moreover, the power consumption associated with the read and write operations of the RAM is approximately 4000 uW throughout the process. This indicates that proposed framework enables on-chip incremental learning to be achieved with extremely low chip power consumption.}

\section{Conclusion}

Drawing inspiration from the bio-plasticity of dendritic spines, this paper introduces a neuromorphic-friendly, SNN-based incremental learning algorithm called Hebbian weight consolidation (HWC). Additionally, we propose the Neuromorphic on-chip learning framework called mentor-learner on-chip learning framework (MLoC). Experimental results demonstrate that our HWC algorithm 23.19\% outperforms lower bound that without incremental learning algorithm, especially in challenging monkey behavior decoding scenarios. Considering on-chip computing, our proposed algorithm exhibits an improvement of 11.06\%. In the behavior decoding scenario, our off-chip method surpasses other SoTA incremental learning algorithms with improvements ranging from 4.88\% to 10.43\%. These results highlight the notable performance of the HWC algorithm in difficult tasks, particularly in the classification of long sequences where catastrophic forgetting is inevitable, demonstrating its ability to significantly delay the forgetting process.

Furthermore, the MLoC method presents a memristor-free on-chip learning framework that does not rely on highly customized chip architectures. It enables the deployment of various known incremental learning algorithms on neuromorphic chips for incremental learning tasks. By implementing the MLoC framework on Neuromorphic chips, we tap into the enormous potential of these chips for incremental learning, including applications in Brain-Machine Interfaces (BMI) hardware. The combination of the HWC algorithm and the on-chip learning framework on Neuromorphic chips demonstrates significant potential for achieving high-performance incremental learning. It addresses challenges in long sequence classification, avoids catastrophic forgetting, and unlocks the capabilities of BMI hardware.

%Bibliography
\bibliographystyle{unsrt}  
\bibliography{references}

\begin{thebibliography}{10}

\bibitem{liu2020neural}
Zhengwu Liu, Jianshi Tang, Bin Gao, Peng Yao, Xinyi Li, Dingkun Liu, Ying Zhou, He~Qian, Bo~Hong, and Huaqiang Wu.
\newblock Neural signal analysis with memristor arrays towards high-efficiency brain--machine interfaces.
\newblock {\em Nature communications}, 11(1):4234, 2020.

\bibitem{gallego2020long}
Juan~A Gallego, Matthew~G Perich, Raeed~H Chowdhury, Sara~A Solla, and Lee~E Miller.
\newblock Long-term stability of cortical population dynamics underlying consistent behavior.
\newblock {\em Nature neuroscience}, 23(2):260--270, 2020.

\bibitem{ma2022neuromorphic}
Songchen Ma, Jing Pei, Weihao Zhang, Guanrui Wang, Dahu Feng, Fangwen Yu, Chenhang Song, Huanyu Qu, Cheng Ma, Mingsheng Lu, et~al.
\newblock Neuromorphic computing chip with spatiotemporal elasticity for multi-intelligent-tasking robots.
\newblock {\em Science Robotics}, 7(67):eabk2948, 2022.

\bibitem{zhang2022recent}
Duzhen Zhang, Shuncheng Jia, and Qingyu Wang.
\newblock Recent advances and new frontiers in spiking neural networks.
\newblock {\em arXiv preprint arXiv:2204.07050}, 2022.

\bibitem{wang2023complex}
Qingyu Wang, Tielin Zhang, Minglun Han, Yi~Wang, Duzhen Zhang, and Bo~Xu.
\newblock Complex dynamic neurons improved spiking transformer network for efficient automatic speech recognition.
\newblock In {\em Proceedings of the AAAI Conference on Artificial Intelligence}, volume~37, pages 102--109, 2023.

\bibitem{cheng2023meta}
Xiang Cheng, Tielin Zhang, Shuncheng Jia, and Bo~Xu.
\newblock Meta neurons improve spiking neural networks for efficient spatio-temporal learning.
\newblock {\em Neurocomputing}, 531:217--225, 2023.

\bibitem{jiang2023origin}
Zhiwei Jiang, Jiaming Xu, Tielin Zhang, Mu-ming Poo, and Bo~Xu.
\newblock Origin of the efficiency of spike timing-based neural computation for processing temporal information.
\newblock {\em Neural Networks}, 160:84--96, 2023.

\bibitem{roy2019towards}
Kaushik Roy, Akhilesh Jaiswal, and Priyadarshini Panda.
\newblock Towards spike-based machine intelligence with neuromorphic computing.
\newblock {\em Nature}, 575(7784):607--617, 2019.

\bibitem{singh2021rectification}
Pravendra Singh, Pratik Mazumder, Piyush Rai, and Vinay~P Namboodiri.
\newblock Rectification-based knowledge retention for continual learning.
\newblock In {\em Proceedings of the IEEE/CVF conference on computer vision and pattern recognition}, pages 15282--15291, 2021.

\bibitem{wu2019large}
Yue Wu, Yinpeng Chen, Lijuan Wang, Yuancheng Ye, Zicheng Liu, Yandong Guo, and Yun Fu.
\newblock Large scale incremental learning.
\newblock In {\em Proceedings of the IEEE/CVF conference on computer vision and pattern recognition}, pages 374--382, 2019.

\bibitem{tang2023neuro}
Yushun Tang, Ce~Zhang, Heng Xu, Shuoshuo Chen, Jie Cheng, Luziwei Leng, Qinghai Guo, and Zhihai He.
\newblock Neuro-modulated hebbian learning for fully test-time adaptation.
\newblock In {\em Proceedings of the IEEE/CVF Conference on Computer Vision and Pattern Recognition}, pages 3728--3738, 2023.

\bibitem{jia2023explaining}
Shuncheng Jia, Tielin Zhang, Ruichen Zuo, and Bo~Xu.
\newblock Explaining cocktail party effect and mcgurk effect with a spiking neural network improved by motif-topology.
\newblock {\em Frontiers in Neuroscience}, 17:1132269, 2023.

\bibitem{zhang2022multi}
Duzhen Zhang, Tielin Zhang, Shuncheng Jia, and Bo~Xu.
\newblock Multi-sacle dynamic coding improved spiking actor network for reinforcement learning.
\newblock In {\em Proceedings of the AAAI Conference on Artificial Intelligence}, volume~36, pages 59--67, 2022.

\bibitem{krotov2023new}
Dmitry Krotov.
\newblock A new frontier for hopfield networks.
\newblock {\em Nature Reviews Physics}, pages 1--2, 2023.

\bibitem{mishra2023survey}
Richa Mishra and Manan Suri.
\newblock A survey and perspective on neuromorphic continual learning systems.
\newblock {\em Frontiers in Neuroscience}, 17:1149410, 2023.

\bibitem{chakraborty2020pathways}
Indranil Chakraborty, A~Jaiswal, AK~Saha, SK~Gupta, and K~Roy.
\newblock Pathways to efficient neuromorphic computing with non-volatile memory technologies.
\newblock {\em Applied Physics Reviews}, 7(2), 2020.

\bibitem{prezioso2018spike}
M~Prezioso, MR~Mahmoodi, F~Merrikh Bayat, H~Nili, H~Kim, A~Vincent, and DB~Strukov.
\newblock Spike-timing-dependent plasticity learning of coincidence detection with passively integrated memristive circuits.
\newblock {\em Nature communications}, 9(1):5311, 2018.

\bibitem{wu2021atomically}
Liangmei Wu, Aiwei Wang, Jinan Shi, Jiahao Yan, Zhang Zhou, Ce~Bian, Jiajun Ma, Ruisong Ma, Hongtao Liu, Jiancui Chen, et~al.
\newblock Atomically sharp interface enabled ultrahigh-speed non-volatile memory devices.
\newblock {\em Nature nanotechnology}, 16(8):882--887, 2021.

\bibitem{kirkpatrick2017overcoming}
James Kirkpatrick, Razvan Pascanu, Neil Rabinowitz, Joel Veness, Guillaume Desjardins, Andrei~A Rusu, Kieran Milan, John Quan, Tiago Ramalho, Agnieszka Grabska-Barwinska, et~al.
\newblock Overcoming catastrophic forgetting in neural networks.
\newblock {\em Proceedings of the national academy of sciences}, 114(13):3521--3526, 2017.

\bibitem{li2017learning}
Zhizhong Li and Derek Hoiem.
\newblock Learning without forgetting.
\newblock {\em IEEE transactions on pattern analysis and machine intelligence}, 40(12):2935--2947, 2017.

\bibitem{zenke2017continual}
Friedemann Zenke, Ben Poole, and Surya Ganguli.
\newblock Continual learning through synaptic intelligence.
\newblock In {\em International conference on machine learning}, pages 3987--3995. PMLR, 2017.

\bibitem{chiavazza2023low}
Stefano Chiavazza, Svea~Marie Meyer, and Yulia Sandamirskaya.
\newblock Low-latency monocular depth estimation using event timing on neuromorphic hardware.
\newblock In {\em Proceedings of the IEEE/CVF Conference on Computer Vision and Pattern Recognition}, pages 4070--4079, 2023.

\bibitem{shanechi2019brain}
Maryam~M Shanechi.
\newblock Brain--machine interfaces from motor to mood.
\newblock {\em Nature neuroscience}, 22(10):1554--1564, 2019.

\bibitem{feulner2022small}
Barbara Feulner, Matthew~G Perich, Raeed~H Chowdhury, Lee~E Miller, Juan~A Gallego, and Claudia Clopath.
\newblock Small, correlated changes in synaptic connectivity may facilitate rapid motor learning.
\newblock {\em Nature communications}, 13(1):5163, 2022.

\bibitem{van2022three}
Gido~M van~de Ven, Tinne Tuytelaars, and Andreas~S Tolias.
\newblock Three types of incremental learning.
\newblock {\em Nature Machine Intelligence}, 4(12):1185--1197, 2022.

\bibitem{rebuffi2017icarl}
Sylvestre-Alvise Rebuffi, Alexander Kolesnikov, Georg Sperl, and Christoph~H Lampert.
\newblock icarl: Incremental classifier and representation learning.
\newblock In {\em Proceedings of the IEEE conference on Computer Vision and Pattern Recognition}, pages 2001--2010, 2017.

\bibitem{van2020brain}
Gido~M Van~de Ven, Hava~T Siegelmann, and Andreas~S Tolias.
\newblock Brain-inspired replay for continual learning with artificial neural networks.
\newblock {\em Nature communications}, 11(1):4069, 2020.

\bibitem{xue2022meta}
Mengqi Xue, Haofei Zhang, Jie Song, and Mingli Song.
\newblock Meta-attention for vit-backed continual learning.
\newblock In {\em Proceedings of the IEEE/CVF Conference on Computer Vision and Pattern Recognition}, pages 150--159, 2022.

\bibitem{wang2022learning}
Zifeng Wang, Zizhao Zhang, Chen-Yu Lee, Han Zhang, Ruoxi Sun, Xiaoqi Ren, Guolong Su, Vincent Perot, Jennifer Dy, and Tomas Pfister.
\newblock Learning to prompt for continual learning.
\newblock In {\em Proceedings of the IEEE/CVF Conference on Computer Vision and Pattern Recognition}, pages 139--149, 2022.

\bibitem{simon2022generalizing}
Christian Simon, Masoud Faraki, Yi-Hsuan Tsai, Xiang Yu, Samuel Schulter, Yumin Suh, Mehrtash Harandi, and Manmohan Chandraker.
\newblock On generalizing beyond domains in cross-domain continual learning.
\newblock In {\em Proceedings of the IEEE/CVF Conference on Computer Vision and Pattern Recognition}, pages 9265--9274, 2022.

\bibitem{zhuang2023gkeal}
Huiping Zhuang, Zhenyu Weng, Run He, Zhiping Lin, and Ziqian Zeng.
\newblock Gkeal: Gaussian kernel embedded analytic learning for few-shot class incremental task.
\newblock In {\em Proceedings of the IEEE/CVF Conference on Computer Vision and Pattern Recognition}, pages 7746--7755, 2023.

\bibitem{wang2022meta}
Zhenyi Wang, Li~Shen, Le~Fang, Qiuling Suo, Donglin Zhan, Tiehang Duan, and Mingchen Gao.
\newblock Meta-learning with less forgetting on large-scale non-stationary task distributions.
\newblock In {\em European Conference on Computer Vision}, pages 221--238. Springer, 2022.

\bibitem{zhou2022forward}
Da-Wei Zhou, Fu-Yun Wang, Han-Jia Ye, Liang Ma, Shiliang Pu, and De-Chuan Zhan.
\newblock Forward compatible few-shot class-incremental learning.
\newblock In {\em Proceedings of the IEEE/CVF conference on computer vision and pattern recognition}, pages 9046--9056, 2022.

\bibitem{zhu2022self}
Kai Zhu, Wei Zhai, Yang Cao, Jiebo Luo, and Zheng-Jun Zha.
\newblock Self-sustaining representation expansion for non-exemplar class-incremental learning.
\newblock In {\em Proceedings of the IEEE/CVF Conference on Computer Vision and Pattern Recognition}, pages 9296--9305, 2022.

\bibitem{shi2015development}
Luping Shi, Jing Pei, Ning Deng, Dong Wang, Lei Deng, Yu~Wang, Youhui Zhang, Feng Chen, Mingguo Zhao, Sen Song, et~al.
\newblock Development of a neuromorphic computing system.
\newblock In {\em 2015 IEEE international electron devices meeting (IEDM)}, pages 4--3. IEEE, 2015.

\bibitem{davies2018loihi}
Mike Davies, Narayan Srinivasa, Tsung-Han Lin, Gautham Chinya, Yongqiang Cao, Sri~Harsha Choday, Georgios Dimou, Prasad Joshi, Nabil Imam, Shweta Jain, et~al.
\newblock Loihi: A neuromorphic manycore processor with on-chip learning.
\newblock {\em Ieee Micro}, 38(1):82--99, 2018.

\bibitem{sandamirskaya2022neuromorphic}
Yulia Sandamirskaya, Mohsen Kaboli, Jorg Conradt, and Tansu Celikel.
\newblock Neuromorphic computing hardware and neural architectures for robotics.
\newblock {\em Science Robotics}, 7(67):eabl8419, 2022.

\bibitem{imam2020rapid}
Nabil Imam and Thomas~A Cleland.
\newblock Rapid online learning and robust recall in a neuromorphic olfactory circuit.
\newblock {\em Nature Machine Intelligence}, 2(3):181--191, 2020.

\bibitem{sumi2020mechanism}
Tomonari Sumi and Kouji Harada.
\newblock Mechanism underlying hippocampal long-term potentiation and depression based on competition between endocytosis and exocytosis of ampa receptors.
\newblock {\em Scientific reports}, 10(1):14711, 2020.

\bibitem{miranda2020modeling}
Enrique Miranda, Gianluca Milano, and Carlo Ricciardi.
\newblock Modeling of short-term synaptic plasticity effects in zno nanowire-based memristors using a potentiation-depression rate balance equation.
\newblock {\em IEEE Transactions on Nanotechnology}, 19:609--612, 2020.

\bibitem{zhu2021self}
Kai Zhu, Yang Cao, Wei Zhai, Jie Cheng, and Zheng-Jun Zha.
\newblock Self-promoted prototype refinement for few-shot class-incremental learning.
\newblock In {\em Proceedings of the IEEE/CVF conference on computer vision and pattern recognition}, pages 6801--6810, 2021.

\bibitem{hu2021distilling}
Xinting Hu, Kaihua Tang, Chunyan Miao, Xian-Sheng Hua, and Hanwang Zhang.
\newblock Distilling causal effect of data in class-incremental learning.
\newblock In {\em Proceedings of the IEEE/CVF conference on Computer Vision and Pattern Recognition}, pages 3957--3966, 2021.

\bibitem{zhang2021self}
Tielin Zhang, Xiang Cheng, Shuncheng Jia, Mu-ming Poo, Yi~Zeng, and Bo~Xu.
\newblock Self-backpropagation of synaptic modifications elevates the efficiency of spiking and artificial neural networks.
\newblock {\em Science advances}, 7(43):eabh0146, 2021.

\bibitem{zhao2020glsnn}
Dongcheng Zhao, Yi~Zeng, Tielin Zhang, Mengting Shi, and Feifei Zhao.
\newblock Glsnn: A multi-layer spiking neural network based on global feedback alignment and local stdp plasticity.
\newblock {\em Frontiers in Computational Neuroscience}, 14:576841, 2020.

\bibitem{zhu2021prototype}
Fei Zhu, Xu-Yao Zhang, Chuang Wang, Fei Yin, and Cheng-Lin Liu.
\newblock Prototype augmentation and self-supervision for incremental learning.
\newblock In {\em Proceedings of the IEEE/CVF Conference on Computer Vision and Pattern Recognition}, pages 5871--5880, 2021.

\bibitem{zhang2023brain}
Tielin Zhang, Xiang Cheng, Shuncheng Jia, Chengyu~T Li, Mu-ming Poo, and Bo~Xu.
\newblock A brain-inspired algorithm that mitigates catastrophic forgetting of artificial and spiking neural networks with low computational cost.
\newblock {\em Science Advances}, 9(34):eadi2947, 2023.

\bibitem{masse2018alleviating}
Nicolas~Y Masse, Gregory~D Grant, and David~J Freedman.
\newblock Alleviating catastrophic forgetting using context-dependent gating and synaptic stabilization.
\newblock {\em Proceedings of the National Academy of Sciences}, 115(44):E10467--E10475, 2018.

\bibitem{kingma2013auto}
Diederik~P Kingma and Max Welling.
\newblock Auto-encoding variational bayes.
\newblock {\em arXiv preprint arXiv:1312.6114}, 2013.

\bibitem{krizhevsky2009learning}
Alex Krizhevsky, Geoffrey Hinton, et~al.
\newblock Learning multiple layers of features from tiny images.
\newblock 2009.

\bibitem{amir2017low}
Arnon Amir, Brian Taba, David Berg, Timothy Melano, Jeffrey McKinstry, Carmelo Di~Nolfo, Tapan Nayak, Alexander Andreopoulos, Guillaume Garreau, Marcela Mendoza, et~al.
\newblock A low power, fully event-based gesture recognition system.
\newblock In {\em Proceedings of the IEEE conference on computer vision and pattern recognition}, pages 7243--7252, 2017.

\bibitem{zhang2018sign}
Qingtian Zhang, Huaqiang Wu, Peng Yao, Wenqiang Zhang, Bin Gao, Ning Deng, and He~Qian.
\newblock Sign backpropagation: An on-chip learning algorithm for analog rram neuromorphic computing systems.
\newblock {\em Neural Networks}, 108:217--223, 2018.

\end{thebibliography}

\end{document}